\documentclass[12pt,reprint,amsmath,amssymb]{revtex4-1}
\usepackage{graphicx}
\usepackage{dcolumn}
\usepackage{subfigure}
\usepackage{epstopdf}
\usepackage[bookmarks=true,colorlinks=true,linkcolor=blue,
urlcolor=blue,citecolor=blue]{hyperref}

\begin{document}

\preprint{APS/123-QED}

\title{\bf The effects of three magnons interactions in the magnon-density waves of triangular spin lattices}

\author{ M. Merdan$^1$ and Y. Xian$^2$}
 \affiliation{%
{$^1$Department of Physics, College of Science, University of Babylon, Hillah, Iraq } \\
$^2$School of Physics and Astronomy, The University of Manchester, Manchester M13 9PL, UK}
\date{\today}

\begin{abstract}
We investigate the magnon-density waves proposed as the longitudinal excitations in triangular lattice antiferromagnets by including the cubic and quartic corrections in the large-$s$ expansion. The longitudinal excitation spectra for the two-dimensional (2D) triangular antiferromagnetic model and quasi-one dimensional (quasi-1D) antiferromagnetic materials have been obtained for a general quantum spin number $s$. For the 2D triangular lattice model, we find a significant reduction (about 40 \%) in the energy spectra at the zone boundaries due to both the cubic and quartic corrections. For the quasi-1D antiferromagnets, since the cubic term comes from the very weak couplings on the hexagonal planes, they make very little correction to the energy spectra, whereas the major correction contribution comes from the quartic terms in the couplings along the chains with the numerical values for the energy gaps in good agreement with the experimental results as reported earlier (Ref.~41).
\end{abstract}
\maketitle

\section{Introduction}

Since Haldane \cite{lifshitz1980statistical} predicted difference between the excitations of integer spin and half-odd-integer spin chains, the nature of excitations of quantum Heisenberg antiferromagnets has attracted  both experimental and theoretical attentions. In particular, for the spin-1 chains, the singlet ground state is separated  from the triplet excitation states by an energy gap. This theoretical prediction  has been confirmed experimentally in the quasi-1D spin-1 antiferromagnetic compounds such as CsNiCl$_3$ and RbNiCl$_3$ \cite{PhysRevLett.56.371}. Haldane's conjecture was also supported by some other experiments \cite{PhysRevB.50.9174,PhysRevLett.56.371, PhysRevLett.69.3571,Steiner1987,PhysRevLett.87.017201} and theoretical studies \cite{PhysRevB.49.13235,PhysRevB.46.10854,
PhysRevLett.75.3348,PhysRevB.48.10227,PhysRevLett.62.2313}. Furthermore, a longitudinal excitation has been proposed by Affleck for explanation of a gapped excitation mode observed in very low temperature in these quasi-1D hexagonal antiferromagnetic compounds CsNiCl${}_3$ and RbNiCl${}_3$ which possess N\'eel order at low temperature \cite{Affleck1989,PhysRevB.46.8934}. This longitudinal mode describes the fluctuations of the long-range order parameter and is beyond the spin-wave theory (SWT) which predicts only the transverse spin-wave excitations (magnons).

On the other hand, the triangular-lattice Heisenberg antiferromagnet is the prototype system of geometrically frustrated magnets and has been under intensive investigation for fundamentally different types of ground and excited states \cite{ANDERSON1973,Fazek1974,Kalmeyer1987}. It is now widely accepted that the ground state of the antiferromagnet on a triangle lattice has the long-range noncollinear N\'eel-like order with the $120^{\circ}$ magnetic three-sublattice structure as predicted by various methods \cite{ springerlink:10.1007/BFb0119592,Springer-Verlag.816.135}, including a SWT based on three-sublattices \cite{Huse1,Jolicoeur,Singh1,Miyake1992, Bernu,Azaria, Elstner1,Chubukov1994, Manuel1, Adolfo2000,Mezzacapo2010}. The interaction between spin-wave excitations in antiferromagnetic  materials of collinear spin configuration is depicted by higher-order anharmonicities beginning with the quartic term \cite{PhysRev.102.1217,PhysRev.117.117}. The higher-order anharmonicities of antiferromagnetic systems with noncollinear spin configuration begin with the cubic term which describes the coupling between transverse (one-magnon) and longitudinal (two- magnon) fluctuations \cite{mou2013,zhi2013}, in addition to the quartic term. This cubic term is similar to those that  describe the interaction between one- and two-particle states of phonons in crystals \cite{ziman1960electrons} and excitations in superfluid bosonic systems \cite{lifshitz1980statistical}. In noncollinear antiferromagnets, the cubic term comes from products of the spin operator components $S^z$  and $S^{x,y}$, which are not present in collinear lattices. For the correction in spin wave spectrum, the cubic term has been included in perturbation theory and represents the coupling of the transverse fluctuations in one sublattice to the longitudinal ones in the others \cite{Miyake1985,Miyake1992, mou2013,zhi2013, JPSJ.62.3277,Chubukov1994,PhysRevB.57.5013}.

For a generic quantum spin-$s$ antiferromagnetic Hamiltonian system with a N\'eel order, a microscopic theory of the longitudinal modes has been proposed \cite{Xian2006}. In this theory the longitudinal excitations are identified as the collective modes of the magnon-density waves, and the corresponding wave functions are constructed by employing the magnon-density operator $S^z$ in similar fashion to Feynman's theory on the low-lying excited states of the helium-4 superfluid where the particle density operator is used \cite{Feynman1954}. In our earlier calculations for the quasi-1D hexagonal structures of CsNiCl$_3$ and RbNiCl$_3$ and tetragonal structure of KCuF${}_3$, we find that, after the inclusion of the higher-order contributions from the quartic terms in the large-$s$ expansion, the energy gap values at the magnetic wavevector are in good agreement with experimental results \cite{PhysRevB.87.174434,xian2014}.

Although there is no report of direct experimental observations of longitudinal modes in 2D triangle antiferromagnetic lattices, a theoretical investigation of dynamic structure factors does find some broad peaks in the two-magnon continuum and a massive contribution from the longitudinal fluctuations to the high energy spectral weight, clearly indicating the strong magnon-magnon interactions in the system \cite{mou2013}. In this article, we extend our preliminary investigation of the longitudinal modes in the 2D triangle antiferromagnetic model \cite{M.Merdan2012}, focusing now on the higher-order calculations by including both the cubic and quartic terms. Our results show a significant reduction on the energy spectra due to the high order corrections. We also examine the cubic term contribution to the energy spectrum correction for the quasi-1D hexagonal systems of CsNiCl$_3$ and RbNiCl$_3$, not considered in our earlier study \cite{PhysRevB.87.174434}. We find that in these systems the cubic term contribution is negligible, mainly due to the very weak coupling on the triangular planes of the systems.

We organize this article as follows. Sec.~2 outlines the main results of the spin-wave theory for the triangular lattice model using the bosonization approach. In Sec.~3 and 4 we review our microscopic theory for the longitudinal excitations, including the higher-order corrections from the cubic and quartic terms and using the approximated ground state from SWT, and apply to the 2D triangular antiferromagnetic model. In Sec.~5 we re-examine our calculation of the higher-order corrections in the quasi-1D hexagonal systems where there are several experimental results for comparison. We notice that the energy gap changes very little after inclusion of the cubic contribution, mainly due to the small coefficient for the plane Hamiltonian when compared with the coefficient of the perpendicular (chain) Hamiltonian. In Sec.~6 we conclude this article by a summary and a critical discussion of the longitudinal modes in 2D triangle lattices.

\section{Spin-Wave Formalism for Triangular Lattice Model}

The Heisenberg antiferromagnet on a triangular lattice is described by Hamiltonian with spin operator ${\bf S}$,
\begin{equation}\label{1}
    H=J\sum_{\langle i,j\rangle}{\bf S}_i\cdot {\bf S}_j,
\end{equation}
where $J>0$ is the coupling parameter and the sum on $\langle i,j\rangle$ runs over all the nearest-neighbor pairs of the triangular lattice once. The classical ground state of the antiferromagnetic Heisenberg model on a triangular lattice consists of three sublattices where the direction of each spin on one sublattice forms an angle of $120^\circ$ from those on the other two sublattices. We choose the direction of classical orientation in the $xz$-plane at the  one $i$-sublattice surrounded  by six $j$-sublattices. The Hamiltonian of  Eq.~\eqref{1} can be transformed into a rotating local basis as
\begin{equation}\label{2}
\begin{split}
S_i^x&\rightarrow S_i^x\cos \theta_i+S_i^z\sin\theta_i,\\
S_i^y&\rightarrow S_i^y,\\
S_i^z&\rightarrow S_i^z\cos\theta_i-S_i^x\sin\theta_i,
\end{split}
\end{equation}
where $\theta_i=\mathbf  Q \cdot {\bf r}_i$ and  $\mathbf Q=(4\pi/3,0)$ is the magnetic ordering wavevector of the hexagonal Brillouin zone of the triangular lattice as shown in Fig.~1.
\begin{figure}[h]
\begin{center}
   \includegraphics[scale=0.45]{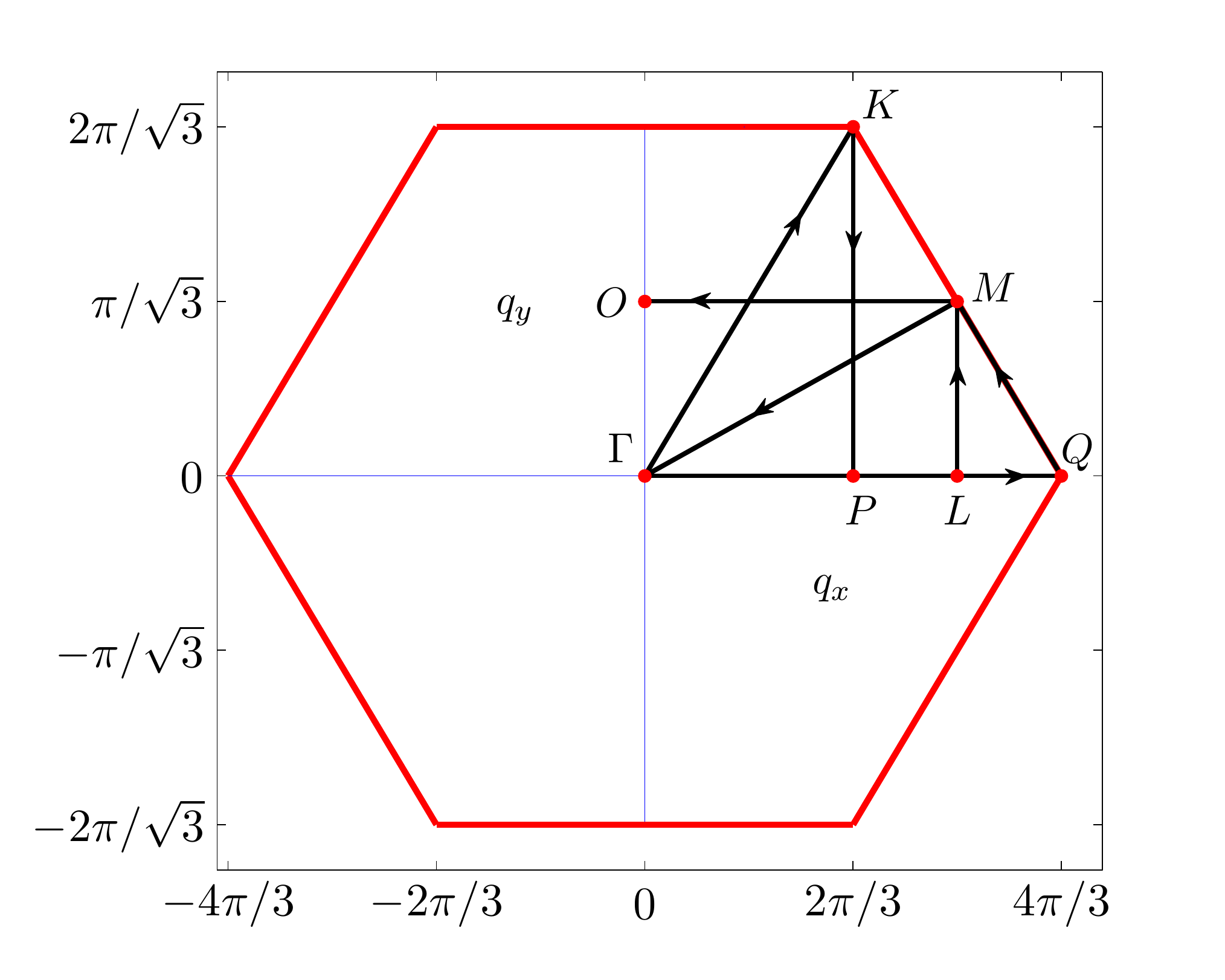}
\caption{\small  The hexagonal first Brillouin zone of a triangular lattice in reciprocal space. The coordinates of the labeled points are, $\Gamma=(0,0)$, $P=(2\pi/3,0)$, $L=(\pi,0)$, $Q=(4\pi/3,0)$, $M=(\pi,\pi/\sqrt3)$, $K=(2\pi/3,2\pi/\sqrt3)$ and $O=(0,\pi/\sqrt3)$.}
\end{center}
\end{figure}
\\
The Hamiltonian operator of Eq.~\eqref{1} after this transformation is given by
\begin{align}\label{3}
    H&=J\sum\limits_{\langle i,j\rangle}\big[ \cos(\theta_i-\theta_j)(S_i^xS_j^x+S_i^zS_j^z)+\xi    S_i^yS_j^y\nonumber\\
    &+\sin(\theta_i-\theta_j)
    (S_i^zS_j^x-S_i^xS_j^z)\big],
\end{align}
where we have also introduced an anisotropy parameter $\xi    (\le1)$ along the $y$-axis. The Holstein-Primakoff transformation which transforms spin operators into bosons is used for the spin-wave calculations such that
\begin{equation}\label{4}
S_i^z = s-a_i^\dagger a_i,\quad
S_i^+ = \sqrt{2s}f_ia_i,\quad S_i^-=\sqrt{2s}a_i^\dagger f_i,
\end{equation}
where $f_i=\sqrt{1-a_i^\dagger a_i/2s}$, $s$ is the spin quantum number and  $S_i^{\pm}=S_i^x\pm iS_i^y$. Substituting Eq.~\eqref{4} into Eq.~\eqref{3} and approximating the expansion of the square root in $f_i$  to the first order in $a_i^\dagger a_i/2s$, we obtain the following Hamiltonian
\begin{equation}\label{5}
H=H_0+H_1+H_2+H_3+H_4,
\end{equation}
where $H_0=-3/2JNs^2$  is the classical ground-state energy ${\cal O}(s^2)$, $H_2$ is the harmonic part of the linear SWT (LSWT) correction ${\cal O}(s)$, $H_3$ is the cubic anharmonic term ${\cal O}(s^{1/2})$ and $H_4$ is the quartic anharmonic term ${\cal O}(s^{0})$. The LSWT  depicts the harmonic approximation or noninteracting magnons. The quadratic terms in $H_2$ can be written as
\begin{align}\label{eq6}
H_2=\frac{1}{4}Js&\sum\limits_{\langle i,j\rangle}\big[2(n_i+n_j)
-(1+2\xi   )(a_ia_j+a_i^{\dagger}a_j^{\dagger}) \nonumber \\
&-(1-2\xi   )(a_ia_j^{\dagger}+a_i^{\dagger}a_j)\big],
\end{align}
where $n_i=a_i^\dagger a_i$ and $n_j=a_j^\dagger a_j$ are number operators. After Fourier transformation for the boson operators with the Fourier component operators $a_{\bf{q}}$ and $a_{\bf{q}}^{\dagger}$, and performing the diagonalization of $H_2$ by the canonical Bogoliubov transformation, $a_{\bf{q}}=u_{\bf{q}}\alpha_{\bf{q}}+v_{\bf{q}}\alpha_{-{\bf{q}}}
^\dagger$, the linear spin-wave Hamiltonian now reads
\begin{equation}\label{7}
   H'=H_0+H_2=-\frac{3}{2}JNs(s+1)+\sum_{\bf{q}}{\cal E}_{\bf{q}}(\alpha_{\bf{q}}^\dagger \alpha_{\bf{q}}+\frac{1}{2}),
\end{equation}
where ${\cal E}_{\bf{q}}=3Js\,\omega_{\bf{q}}$ is the spin-wave excitation spectrum with the dimensionless spectrum $\omega_{\bf{q}}$ given by
\begin{equation}\label{8}
\omega_{\bf{q}}=\sqrt{A_{\bf{q}}^2-B_{\bf{q}}^2}=\sqrt{(1-\gamma_{\bf{q}})(1+2\xi   \gamma_{\bf{q}})}\,,
\end{equation}
with $A_{\bf{q}}$ and $B_{\bf{q}}$ defined by
\begin{equation}\label{9}
    A_{\bf{q}}=1+\big(\xi-\frac{1}{2}\big)\gamma_{\bf{q}},\quad
    B_{\bf{q}}=\big(\xi+\frac{1}{2}\big)\gamma_{\bf{q}},
\end{equation}
respectively, and $\gamma_{\bf{q}}$ defined by
\begin{equation}\label{10}
    \gamma_{\bf{q}}=\frac{1}{z}\sum_\varrho e^{i\mathbf {q\cdot r}_\varrho}=\frac{1}{3}\big(\cos q_x+2\cos
    \frac{q_x}{2}\cos\frac{\sqrt{3}}{2}q_y\big),
\end{equation}
with the summation over the nearest-neighbor index $\varrho$ and the coordination number $z=6$ for the triangular lattice.

The cubic term exists in the triangular lattice because the coupling of $S^z$ and $S^x$ spin components. In terms of boson operators the cubic term reads
\begin{equation}\label{11}
    H_3=J\sqrt{\frac{s}{2}}\sum\limits_{\langle i,j\rangle}\sin(\theta_i-\theta_j)\big[
    (a_i+a_i^{\dagger})n_j-n_i(a_j+a_j^{\dagger})\big].
\end{equation}
We notice that for the collinear spin lattices, $\sin(\theta_i-\theta_j)=0$ and the cubic terms vanish and that $H_1$ with one boson terms always cancel out. Furthermore, the LSWT ground-state expectation value of the three-boson operators is always zero. This cubic term has been included in the perturbation theory with the  contribution of order $1/s^2$. In more details, after performing  Fourier and Bogoliubov transformations, we obtain
\begin{align}\label{12}
H_3=Jzi\sqrt{\frac{3s}{8N}}&\sum\limits_{\bf q,k}\Big[\frac{1}{2!}{ F}_1(\bf{q,k})\alpha_{\bf q}^\dagger \alpha_{\bf k-q}^\dagger\alpha_{{\bf k}} \nonumber \\
&+\frac{1}{3!}{ F}_2({\bf{q,k}})\alpha_{{\bf q}}^\dagger  \alpha_{{\bf k+q}}^\dagger\alpha_{{\bf k}}^\dagger+\text{H.c.}\Big],
\end{align}
with ${ F}_1(\bf{q,k})$ and ${ F}_2(\bf{q,k})$ given by
\begin{align}\label{13}
{F_1}({\mathbf{q}},{\mathbf{k}}) = &{\bar \gamma _{\mathbf{q}}}({u_{\mathbf{q}}} + {v_{\mathbf{q}}})({u_{\mathbf{k}}}{u_{{\mathbf{q}} - {\mathbf{k}}}} + {v_{\mathbf{k}}}{v_{{\mathbf{q}} - {\mathbf{k}}}}) \nonumber\\
&+ {\bar \gamma _{\mathbf{k}}}({u_{\mathbf{k}}} + {v_{\mathbf{k}}})({u_{\mathbf{q}}}{u_{{\mathbf{q}} - {\mathbf{k}}}} + {v_{\mathbf{q}}}{v_{{\mathbf{q}} - {\mathbf{k}}}})\nonumber\\
 &- {\bar \gamma _{{\mathbf{q}} - {\mathbf{k}}}}({u_{{\mathbf{q}} - {\mathbf{k}}}} + {v_{{\mathbf{q}} - {\mathbf{k}}}})({u_{\mathbf{q}}}{u_{\mathbf{k}}} + {v_{\mathbf{q}}}{v_{\mathbf{k}}}),
\end{align}
and
\begin{align}\label{14}
{F_2}({\mathbf{q}},{\mathbf{k}}) =& {\bar \gamma _{\mathbf{q}}}({u_{\mathbf{q}}} + {v_{\mathbf{q}}})({u_{\mathbf{k}}}{v_{{\mathbf{q}} + {\mathbf{k}}}} + {v_{\mathbf{k}}}{u_{{\mathbf{q}} + {\mathbf{k}}}}) \nonumber\\
&  + {\bar \gamma _{\mathbf{k}}}({u_{\mathbf{k}}} + {v_{\mathbf{k}}})({u_{\mathbf{q}}}{v_{{\mathbf{q}} + {\mathbf{k}}}} + {v_{\mathbf{q}}}{u_{{\mathbf{q}} + {\mathbf{k}}}})\nonumber\\
&- {\bar \gamma _{{\mathbf{q}} + {\mathbf{k}}}}({u_{{\mathbf{q}} + {\mathbf{k}}}} + {v_{{\mathbf{q}} + {\mathbf{k}}}})({u_{\mathbf{q}}}{v_{\mathbf{k}}} + {v_{\mathbf{q}}}{u_{\mathbf{k}}}),
\end{align}
where $u_i$ and $v_i$ are Bogoliubov parameters and the function $\bar \gamma _{\mathbf{q}}$ is given by
\begin{equation}
  \bar \gamma _{\mathbf{q}}=\frac{1}{3}\big(\sin q_x-2\sin \frac{q_x}{2}\cos\frac{\sqrt{3}}{2}q_y\big).
\end{equation}
The first term in Eq.~\eqref{12} is called "decay" which describes the interaction between one- and two-magnon states, and it is symmetric under permutation of two outgoing momenta. The second term is called "source", and it is symmetric under permutation of three outgoing momenta \cite{Chernyshev2009a}.
The $1/s^2$ contribution from the second-order perturbation of $H_3$ is evaluated by Miyake \cite{Miyake1985,Miyake1992} such that
\begin{equation}\label{16}
  \delta E_3=-\frac{z^2J^2s}{16N}\sum_{\bf q,k}\frac{ F_2({\bf{q,k}})^2}{{\cal E}_{\bf{q}}+{\cal E}_{\bf{k}}+{\cal E}_{\bf{q+k}}}.
\end{equation}
The quartic anharmonic term in Eq.~\eqref{5} reads
\begin{align}\label{18}
H_4=\frac{1}{4}J&\sum\limits_{\langle i,j\rangle}\Big[-n_in_j
+\frac14(1+2\xi   )(n_i+n_j)a_ia_j\nonumber\\
&+(1-2\xi   )\{a_j^{\dagger}(n_i+n_j)a_i+\text{H.c.}\Big].
\end{align}
For simplicity, we define the following Hartree-Fock averages (the LSWT ground-state expectation values) of the triangular lattice
\begin{equation}\label{18a}
\begin{split}
&\rho=\langle a_l^\dagger a_l\rangle=\frac{1}{N}\sum_{\bf q}\rho_{\bf q},\quad\mu_\varrho=\langle a_l^\dagger a_{l+\varrho}\rangle=\frac{1}{N}\sum_{\bf q}e^{i\bf {\bf q}\cdot\varrho}\rho_{\bf q}, \\
& \Delta_\varrho=\langle a_l a_{l+\varrho}\rangle=\frac{1}{N}\sum_{\bf q}e^{i\bf {\bf q}\cdot\varrho}\Delta_{\bf q},\quad\delta=\langle a_l a_l\rangle=\frac{1}{N}\sum_{\bf q}\Delta_{\bf q},
\end{split}
\end{equation}
with $\Delta_{\bf q}$ and $\rho_{\bf q}$ defined as
\begin{equation}\label{19}
\Delta_{\bf q}=\frac{1}{2}\frac{B_{\bf q}}{\sqrt{A_{\bf q}^2-B_{\bf q}^2}},\quad \rho_{\bf q}=\frac{1}{2}\big(\frac{A_{\bf q}}{\sqrt{A_{\bf q}^2-B_{\bf q}^2}}
    -1\big).
\end{equation}
The ground-state expectation value of the quartic $H_4$ of Eq.~\eqref{18} can be calculated first by applying Fourier transformation and then Bogoliubov transformation using Wick's theorem. The ground state energy correction in terms of the Hartree-Fock averages is given by
\begin{align}
\delta E_4=-\frac{1}{4}JNz&\Big[\rho^2+\mu_\varrho^2+\Delta_\varrho^2-(1+2\xi   )(\rho\Delta_\varrho+\frac12\mu_\varrho\delta) \nonumber\\
&-(1-2\xi   )(\rho\mu_\varrho+\frac12\Delta_\varrho\delta)\Big].
\end{align}
Thus, the total ground state energy can be calculated from all these contributions for the isotropic case $\xi=1$ as
\begin{align}
 E=-\frac{1}{4}JNzs^2&\Big[1+\frac{I_2}{s}+\frac{(I_3+I_4)}{(2s)^2}\Big],
\end{align}
where $I_2$ is related to harmonic part $H_2$ with numerical value given by
\begin{equation}
I_2=1-\frac1N\sum_{\bf q} \omega_{\bf q}=0.218412,
\end{equation}
and the other constants $I_3$ and $I_4$ are related to $\delta E_3$ and $\delta E_4$ respectively with numerical values calculated at $\xi   =1$
\begin{equation}\label{23}
  I_3=\frac{2}{N^2}\sum_{\bf q,k}\frac{ F_2({\bf{q,k}})^2}{{\omega}_{\bf{q}}
  +{\omega}_{\bf{k}}+{\omega}_{\bf{q+k}}}=0.2756(2),
\end{equation}
\begin{align}
  I_4=4\Big(\rho^2&+\mu_\varrho^2+\Delta_\varrho^2-3
  (\rho\Delta_\varrho+\frac12\mu_\varrho\delta) \nonumber\\
&+(\rho\mu_\varrho+\frac12\Delta_\varrho\delta)\Big)=-0.25429.
\end{align}
These numerical results have been obtained by Miyake \cite{Miyake1992}. The integration of $I_3$ is four dimensional integral and  has been calculated  by Monte Carlo integration using Mathematica software.

The sublattice magnetization $M$ in general can be written in terms of the magnon density $\rho$ as
\begin{equation}\label{25}
M=s-\rho.
\end{equation}
Within the linear spin-wave approximation, $\rho$ is given by $\rho_0=0.261303$. The higher-order correction to the sublattice magnetization can be expressed as $M=s-\rho_0+\frac{\delta s_2}{2s}$.
Miyake first calculated $\delta s_2=0.0110$ \cite{Miyake1992},  later confirmed by Chernyshev and Zhitomirsky \cite{Chernyshev2009a} using a different method. But Chubukov \cite{Chubukov1994} obtained $\delta s_2=0.027$, perhaps because of an integration problem.

\section{Longitudinal Excitations Formalism}

In antiferromagnetic quantum systems with a N\'eel-like long-range order, the longitudinal excitations correspond to the fluctuations in the order parameter.  We identify the longitudinal modes as the magnon-density waves (MDW), well defined only in the systems where the interactions between the transverse magnons are significant. In the low dimensional systems the magnon density is high enough to support the longitudinal waves, such as the cases of the quasi-1D systems mentioned in Sec.~1. It remains questionable whether or not the interaction between magnons in pure 2D systems such as the triangle antiferromagnet is strong enough to support the longitudinal modes, although there is some indication this may be so \cite{mou2013}.

The magnon density operator is given by spin operator $S^z$ and so $S^z$ can be used to construct the wave function of longitudinal excitation state in similar fashion as Feynman's theory of the phonon-roton excitation state of the helium superfluid, where the density operator is the usual particle density operator \cite{Feynman1954, Feynman1956}.
The longitudinal excitation state is thus constructed by applying the magnon density fluctuation operator $X_{\bf q}$ on the ground state $|\Psi_g\rangle$ as
\begin{equation}\label{26}
|\Psi_e\rangle =X_{\bf q}|\Psi_g\rangle,
\end{equation}
where $X_{\bf q}$ is given in terms of the Fourier transformation of $S^z$ operator as,
\begin{equation}\label{27}
X_{\bf q} = \frac{1}{\sqrt{N}}\sum_l e^{i\mathbf {q\cdot r}_l} S^z_l,\quad q>0,
\end{equation}
with index $l$ running over all lattice sites. The condition $q>0$ in Eq.~\eqref{27} guarantees the orthogonality to the ground state. The energy spectrum of longitudinal excitation is given by \cite{Xian2007}
\begin{equation}\label{28}
    E({\bf q})=\frac{N({\bf q})}{S({\bf q})},
\end{equation}
where $N({\bf q})$ is  the ground-state expectation value of a double commutator such that
\begin{equation}\label{29}
    N({\bf q})=\frac{1}{2}\langle[X_{-{\bf q}},[H,X_{\bf q}]]\rangle_g,
\end{equation}
and $S({\bf q})$ is the normalization integral or the structure factor of the lattice model
\begin{equation}\label{30}
    S({\bf q})=\langle\Psi_e|\Psi_e\rangle=\frac{1}{N}\sum_{l,l'}e^{i\mathbf q.(\mathbf r_{l}-\mathbf r_{l'})}\langle S_l^zS_{l'}^z\rangle_g.
\end{equation}
We apply the SWT for the approximation of the ground state $|\Psi_g\rangle$ in the following sections to calculate these expectation values.

\section{Magnon-density waves in 2D Triangular lattice}

The one-sublattice Hamiltonian of Eq.~\eqref{3}  after the rotation for the triangular lattice is used to obtain the double commutator of Eq.~\eqref{29} as
\begin{equation}\label{31}
   N({\bf q})=\frac{1}{4}sJ\sum_\varrho\Big[(1+2\xi)
   (1+\gamma_q)\tilde g_\varrho+(1-2\xi)
   (1-\gamma_q)\tilde g'_\varrho+\frac{I_3}{8s}\Big],
   \end{equation}
where $I_3$ contains cubic terms and is defined in Eq.~\eqref{23}, and the transverse correlation functions $\tilde g_\varrho$ and $\tilde g'_\varrho$ are defined as
\begin{equation}\label{32}
    \tilde g_\varrho=\frac{1}{2s}\langle S_l^+S_{l+\varrho}^+\rangle_g,\quad   \tilde g'_\varrho=\frac{1}{2s}\langle S_l^+S_{l+\varrho}^-\rangle_g.
\end{equation}
Due to the lattice translational symmetry, both correlation functions are independent of index $l$. These functions contain the contribution from quadratic and quartic terms and both given in terms of the Hartree-Fock averages of Eq.~\eqref{18a} as
\begin{equation}\label{33}
\begin{split}
&\tilde g_\varrho=\Delta_\varrho-\frac{2\rho\,\Delta_\varrho+\mu_\varrho\,\delta}{2s},\\
&\tilde g'_\varrho=\mu_{\varrho}-\frac{2\rho \,\mu_{\varrho}+\Delta_{\varrho}\delta}{2s}.
\end{split}
\end{equation}
 We obtain the numerical results at the isotropic point $\xi=1$ as $\tilde g_\varrho=0.12598$ and $\tilde g_\varrho'=0.03642$ for all the six nearest neighbors. As it can be seen, $N(\bf q)$ is dominated by $\tilde g_\varrho$.
The structure factor  is independent of $s$, and is given by
\begin{equation}\label{34}
    S({\bf q})=\rho+\frac{1}{N}\sum_{\bf q'}\rho_{\bf q'}\rho_{\bf q+q'}+\frac{1}{N}
    \sum_{\bf q'}\Delta_{\bf q'}\Delta_{\bf q+q'}.
\end{equation}
We notice that the calculations of both Eqs.~\eqref{33} and \eqref{34} involve up to four-boson operators of the quartic terms, but not the cubic term. We then calculate the longitudinal excitation spectrum $E({\bf q})$ given by Eq.~\eqref{28}. From the numerical calculation, we found that this spectrum of the longitudinal mode is gapless in the thermodynamic limit, as $E({\bf q})\to 0$ at both $ {\bf q}\to0$ and ${\mathbf q}\to\pm \mathbf Q$. Two longitudinal modes for the triangular lattice antiferromagnets due to the noncollinear nature of the N\'eel-like order can be obtained by folding of the wavevector. We denote one as $L_+$ with the spectrum $E(\bf q+Q)$ and the other as $L_-$ with the spectrum $E(\bf q-Q)$. We plot both spectra at isotropic case $\xi=1$ in Figs.~\ref{2} and \ref{3}. We find that the energy values of the spectra reduce significantly by about 40\% at the zone boundaries after inclusion of the quartic and cubic corrections, and that the two longitudinal modes are nearly degenerate, only differ by a few percents on the zone boundaries. For example, the $L_+$ energy value at $P$ is $0.6545zsJ$ in the first order calculation, reduces to $0.3829zsJ$ after including the higher-order corrections with the cubic contribution of $0.0933zsJ$ and the quartic contribution of $-0.3648zsJ$. The spectrum for both modes is still gapless at $\Gamma$ point where $\gamma_{\bf q}=1$, and at the points $K$ and $Q$ where $\gamma_{\bf q}=1/2$.

The numerical calculation demonstrates that the gapless spectra of $L_-$ and $L_+$ modes are expected due to the slow logarithmic divergence in both the second and third terms in the structure factor $S({\bf q})$ of Eq.~\eqref{34}. More precisely, near $\Gamma$, $K$ and $Q$ points, we find that $S({\bf q})\propto -\ln q$, and hence the excitation spectrum $E({\bf q})\propto-1/\ln q$ with different coefficients.

 \begin{figure}[h!]
\centering
\includegraphics[scale=0.45]{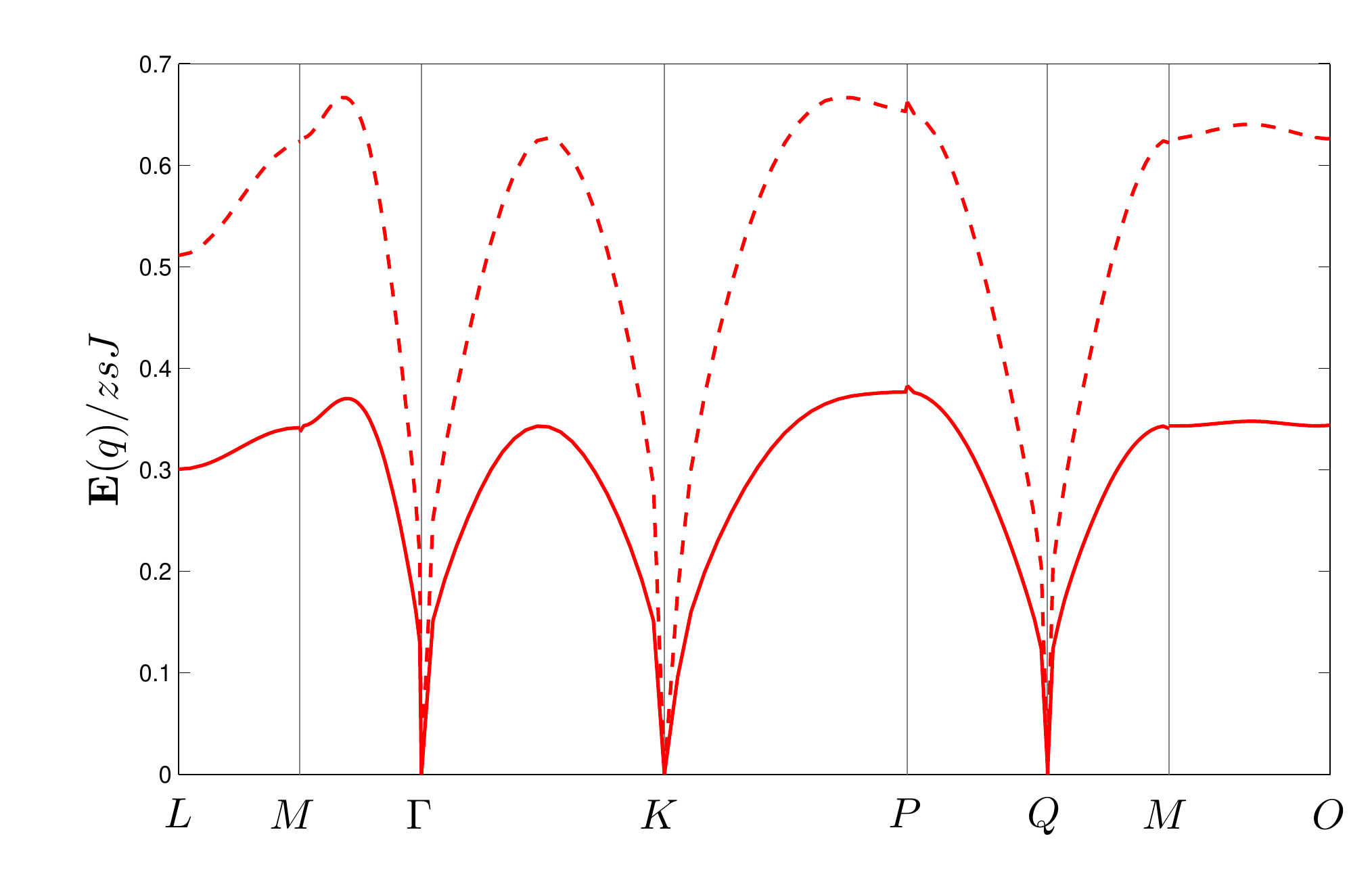}\\
  \caption{The excitation spectrum  of the longitudinal mode $L_+$  along ($LM\Gamma KPQMO$) of the BZ with isotropic case $\xi=1$. It is gapless at $\Gamma$, $K$ and $Q$ points. The longitudinal
spectra calculated from the first-order and higher-order approximations are indicated by the
dash and solid lines respectively.}
  \label{2}
\end{figure}
 \begin{figure}[h!]
\centering
\includegraphics[scale=0.45]{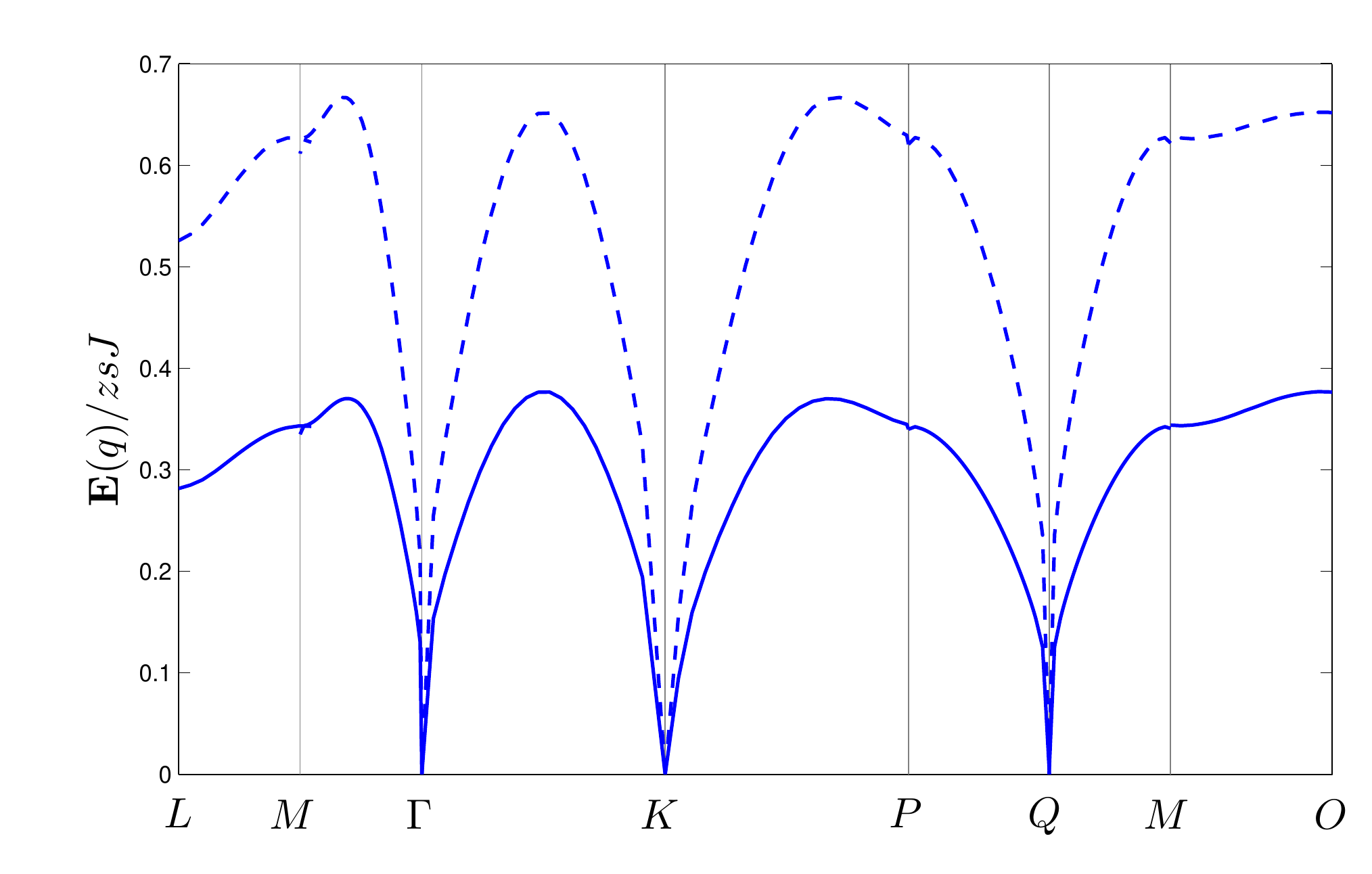}\\
  \caption{The excitation spectrum  of the longitudinal mode $L_-$  along ($LM\Gamma KPQMO$) of the BZ with isotropic case $\xi=1$ . It is gapless at $\Gamma$, $K$ and $Q$ points. The longitudinal
spectra calculated from the first-order, and higher-order approximations are indicated by the
dash and solid lines respectively.}
  \label{3}
\end{figure}

\begin{figure}[h]
\centering
\includegraphics[scale=0.45]{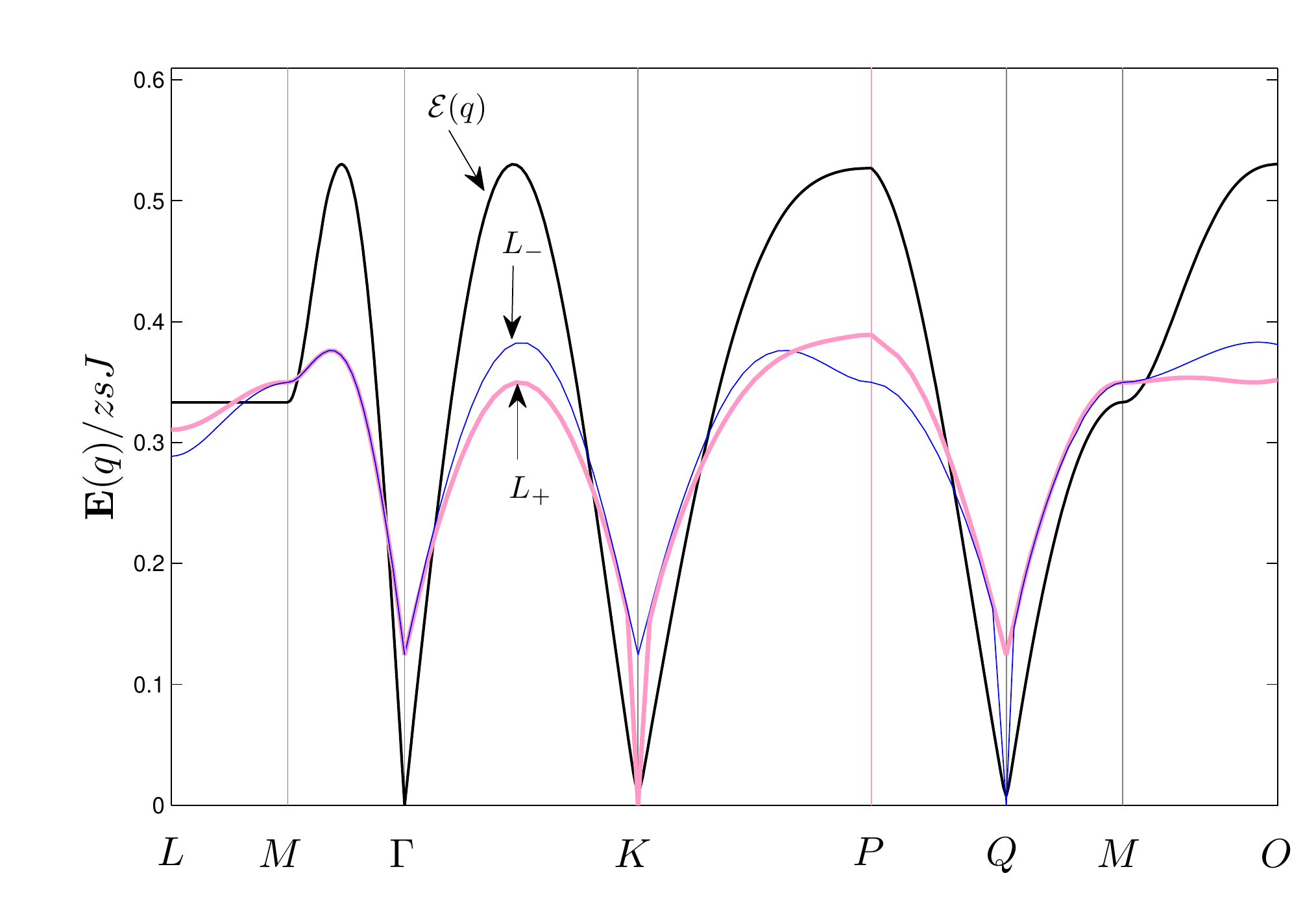}\\
  \caption{The longitudinal modes $L_-$ and $L_+$ together with spin-wave excitation spectrum ${\cal E}(q)$ of the 2D triangle lattice along ($LM\Gamma KPQMO$) of the BZ with an anisotropy $\xi=1-1.5\times10^{-4}$. The longitudinal gap values at $\Gamma$, $K$, and $Q$ points for both modes, $L_-$, and $L_+$ respectively   are $0.2030zsJ$ in the first-order approximation and $0.1242szJ$ after including the high order correction. The transverse spin-wave gap is $0.0075zsJ$.}
  \label{6}
\end{figure}

The logarithmic behavior of the structure factor and  the energy spectrum of the triangular lattice model is similar to that of the square lattice model investigated earlier \cite{Xian2006,Xian2007,yang2011,xian2014}. We have identified these gapless modes of the 2D models as quasi-gapped modes because any finite size effect or anisotropy will induce a large energy gap when compared with the counterparts of the spin-wave spectrum. The effect of anisotropy can be investigated by considering the value of $\xi$ parameter in Eq.~\eqref{3} differing from unity. For example, for a tiny anisotropy such as $\xi=1-1.5\times10^{-4}$, at $\Gamma$ point for both modes, we obtain the energy gap value of $0.2030zsJ$ in the first-order approximation and $0.1242szJ$ after including the high order corrections with the cubic contribution of $0.0407zsJ$ and quartic contribution of $-0.1195zsJ$. The gap value of the corresponding spin-wave spectrum at the same value of anisotropy is $0.0075zsJ$, much smaller. In particular, we find that the longitudinal energy gap value is proportional to $1/[-\ln(1-\xi)]$, in compared with the spin-wave gap which is proportional to $\sqrt{1-\xi}$, when $\xi\rightarrow1$. In order to make further comparison between the longitudinal mode and the transverse spin-waves mode, we plot both the spectra with $\xi=1-1.5\times10^{-4}$ in Fig.~\ref{6} along the path $(LM\Gamma KPQMO)$ of the BZ. The different gap values for the longitudinal and transverse mode at $\Gamma$, $K$ and $Q$ points can be clearly seen. The spin-wave spectra at $\Gamma$ point are still gapless where $\gamma_{\mathbf q}=1$ whereas both longitudinal modes have the gap value of $0.1242szJ$. The $L_+$ mode has the same gap at $Q$ point, but it is gapless at $K$ point, and vice versa for the $L_-$ mode.

Before we turn to the quasi-1D systems in the next section, we like to mention that although we do find the significant energy reduction of the two longitudinal modes after inclusion of the high order terms, we cannot at the moment directly relate our values to the peak structures of dynamic structure calculated in Ref.~32 based on SWT.

\section{Magnon-density waves in qusi-1D triangular lattices}

We now turn to the longitudinal modes for the quasi-1D hexagonal antiferromagnetic systems, modeled by the following Heisenberg Hamiltonian with a strong  interaction $J$ along the chains and weak interaction $J'$ on the hexagonal planes,
\begin{equation}\label{35}
    H=2J\sum_{\langle i,j\rangle}^\text{chain}\mathbf{S}_i\cdot \mathbf{S}_j+2J'\sum_{\langle i,j\rangle}^\text{plane}\mathbf{S}_i\cdot \mathbf{S}_j.
\end{equation}
An energy gap about $0.41(2J)$ has been observed by the neutron scattering experiments for CsNiCl${}_3$ \cite{PhysRevLett.56.371} with spin $s=1$, $J = 0.345$ and $J'=0.0054$ THz. This energy gap does not belong to the transverse spin-wave spectra, but belong to the longitudinal modes, as first proposed by Affleck \cite{Affleck1989,PhysRevB.46.8934}. Following Affleck \cite{Affleck1989,PhysRevB.46.8934}, we earlier calculated the energy gap of the lower longitudinal mode $L_-$ at the point ${\bf Q}=(4\pi/3,0,\pi)$ using Eq.~\eqref{28} but including only the quartic correction and obtained a value of $(0.4907)2J$ \cite{PhysRevB.87.174434}, in reasonable agreement with the experimental result. Now after including the cubic contribution (as described by $I_3$ of Eq.~\eqref{31}), we obtain a value of $(0.4908)2J$ for this gap, with a very small change. For RbNiCl$_3$ also with $s=1$ but $J'/J=0.0295$, the experimental result of of the gap value is about 0.51 THz \cite{PhysRevB.43.13331}, our result is $0.6974$ THz with only quartic correction and $0.6977$ after including cubic correction. The compound CsMnI${}_3$ has spin quantum number $s=5/2$ and the very small ratio of couplings $\epsilon  =J'/J\approx0.005$, for which the SWT approximation for its ground state is very poor, our result for the gap value of $0.47199$ THz with only quartic correction and $0.47200$ THz after including the cubic correction is in very poor comparison with the experimental results of about $0.1$ THz by Harrison \emph{et al} \cite{PhysRevB.43.679}. Clearly in the case of CsMnI${}_3$, we need better ground state in order to obtain better results for the energy gap of the longitudinal modes as mentioned before.

\begin{table}[h!]
\centering
\caption{The numerical results for the energy gap of the $L_-$ mode with and without
the cubic term contribution at the magnetic wavevector for the three quasi-1D materials.}\label{tt}
\setlength{\tabcolsep}{5pt}
\renewcommand{\arraystretch}{1.8}
\vspace{0.1cm}
\begin{tabular}{c|c|c}

  \shortstack{Quasi-1D \\ materials} &\shortstack{$L_-$ mode \\before} & \shortstack{$L_-$ mode \\after}\\
   \hline

  CsNiCl$_3$ & (0.490721)$2J$ & (0.490837)$2J$ \\
  RbNiCl$_3$ & (0.718999)$2J$& (0.719291)$2J$\\
  CsMnI$_3$& (1.19189)$2J$& (1.19191)$2J$\\
  \hline
\end{tabular}
\end{table}

In general, as we can see that from Table.\ref{tt}, the contribution from the cubic term is tiny for the energy spectra of the longitudinal modes. This is mainly due to the small value of  $\epsilon=J'/J$ in these systems, namely the coupling $J'$ on the hexagonal plans with non-vanishing cubic contribution is much smaller than the coupling $J$ along the chains for which cubic term vanishes. The energy spectra of the longitudinal modes for such quasi-1D systems of Eq.~\eqref{35} can be expressed as sum of the chain and plane parts as,
\begin{equation}\label{36}
E_{\pm}=L_{\pm}^{\mathrm{chain}}+L_{\pm}^{\mathrm{plane}}.
 \end{equation}
In Table.\ref{ttt}, we present the numerical results for the energy gap due to the planar term of Eq.~\eqref{36} before and after including cubic corrections for the three quasi-1D materials, and where we define the cubic contribution as $\Delta L_{-}^{\text { plane }}$. We can see that the cubic correction relative to the quartic contribution is similar in ratio to that of the 2D triangular lattice model discussed in Sec.~4.
\begin{table}[h!]
\centering
\caption{The numerical results for the $L_-$energy gap of the  planar term of the Hamiltonian Eq.~\eqref{35} with and without the cubic term contribution at the magnetic
wavevector for the three quasi-1D materials.}\label{ttt}
\renewcommand{\arraystretch}{1.8}
\vspace{0.1cm}
\begin{tabular}{c|c|c|cc}
\shortstack{Quasi-1D \\ materials} &\shortstack{$\,\,\,L_{-}^{\text { plane }}$ \\before} & \shortstack{\,\,\,$L_{-}^{\text { plane }}$ \\after}&\shortstack{$\Delta L_{-}^{\text { plane }}$ \\{}}\\
\cline{1-4}
CsNiCl3 & 0.0309086 $zsJ'$& 0.0333744 $zsJ'$& 0.00246585 $zsJ'$&\\
RbNiCl3& 0.0544659 $zsJ'$    & 0.0577724 $zsJ'$& 0.00330649 $zsJ'$&\\CsMnI3& 0.0231435 $zsJ'$& 0.0237096 $zsJ'$& 0.00056611 $zsJ'$&\\
\cline{1-4}
\end{tabular}
\end{table}

\section{Conclusion}

In this paper, we have investigated the longitudinal excitations of the 2D triangular antiferromagnetic lattice and the quasi-1D hexagonal systems after including the high order corrections. For the 2D triangular model, we find significant reduction of about 40\% in the energy spectra from the higher-order contributions of the cubic and quartic terms. For the quasi-1D hexagonal materials, we find the cubic corrections are negligible when compared with the quartic corrections which was calculated earlier \cite{PhysRevB.87.174434}. This is mainly because of the weak coupling on the triangular planes.

Our numerical values for the energy gap of the longitudinal modes after including the higher-order corrections are in reasonable agreement with the experimental results for the spin-1 compounds CsNiCl$_3$ and RbNiCl$_3$, but is poor for the spin-5/2 compound CsMnI$_3$ because the approximate ground state by SWT is poor for this compound which is very close to a quantum critical point. Clearly a better ground state for this compound will be needed in our calculation of the longitudinal modes in order to make reasonable comparison with the experiment.

Another point that needs addressing is the question of how well defined are the longitudinal modes in 2D triangular antiferromagnets since the magnon density $\rho$ in the order parameter of Eq.~\eqref{25} may not be high enough to support the longitudinal modes. This is similar to the case of the 2D antiferromagnet on a square lattice. As we mentioned earlier, although there is no direct experimental evidence of these longitudinal modes in 2D triangular antiferromagnet, theoretical investigation of dynamic structure factors does find some broad peaks in the two-magnon continuum \cite{mou2013}, indicating the strong longitudinal collective fluctuations. It will also be desirable to investigate the the spontaneous decay of the longitudinal modes due to the coupling to the magnons as represented by the cubic terms in the Hamiltonian \cite{zhi2013} and we wish to report our investigation in the future.

\end{document}